\documentclass[twocolumn,prl,superscriptaddress]{revtex4}

\usepackage{amsmath,amssymb,amsfonts}
\usepackage{CJK}
\usepackage{bm}
\usepackage{array}
\usepackage{multirow}
\usepackage{graphicx}

\begin{document}

\title{Exotic Topological Types of Majorana Zero-modes and Their Universal Quantum Manipulation
}


\author{Y. X. Zhao}
\email[]{zhaoyx@hku.hk}
\author{Z. D. Wang}
\email[]{zwang@hku.hk}

\affiliation{Department of Physics and Center of Theoretical and Computational Physics, The University of Hong Kong, Pokfulam Road, Hong Kong, China}

\date{\today}

\begin{abstract}
From our general index theorem that characterises faithfully the boundary-bulk correspondence of topological superconductors and insulators,  we reveal rigorously that four topologically distinct types of Majorana zero-modes can emerge at the ends of  superconducting wires with various symmetry classes. More intriguingly, we establish three exotic one-dimensional models that have different types of topological charge of Majorana zero-modes, and  disclose exactly the corresponding topological properties, which are respectively different from those of the Kitaev model because their topological essences (i.e., the types of topological charge) are distinct. Moreover, we also address their application in universal quantum manipulation, which is very promising for realising universal topological quantum computation.
\end{abstract}
\maketitle

Recently, the realization of Majorana fermions(MFs) \cite{Majorana,Wilczek},  who are their own anti-particles and usually obey non-Abelian statistics,
has attracted a lot of attention, both theoretically\cite{Kitaev, Kitaev-Honeycomb, Das, Oreg, Cold-atom-I, DIII-I, DIII-II, DIII-III,DIII-IV, BDI-I} and experimentally\cite{Experiment-I, Experiment-II, Experiment-III, Experiment-IV},  because of not only the novel physics but also its potential applications for topological quantum computation\cite{Computation-I, Computation-II, Computation-III, Computation-IV, Exchange-MFs, Review-MFs, RMP}. In particular, topologically protected Majorana zero-modes residing at ends of topological superconductor wires are of special interest, since they are point-like particles topologically stable against disorders and perturbations. The relevant essence was elucidated by Kitaev, i.e., how a nontrivial bulk topological configuration can lead to unpaired Majorana zero-modes at two ends of the superconducting wire\cite{Kitaev, Kitaev-Honeycomb}, not only through illustrating a toy model of spinless p-wave superconductor but also, more importantly, by establishing the topology-related theoretical formalism. Inspired by the Kitaev's pioneering work, various theoretical proposals and experimental endeavours have been made for realising and detecting MFs in one-dimensional (1D) materials~\cite{Das, Oreg, Cold-atom-I, DIII-I, DIII-II, DIII-III, BDI-I, Experiment-I, Experiment-II, Experiment-III, Experiment-IV}, making it very promising to first find MFs in 1D topological superconductors.
However, these proposals are almost focused on concrete 1D systems with a hidden assumption that only one topological type of MFs, as that in the Kitaev model, may exist  even for the 1D models with different symmetry classes. Here we reveal for the first time that four topologically distinct types of Majorana zero-modes can actually emerge at the ends of  1D superconductors, based on a faithful quantitative correspondence in a TI/TSC between the topological charge of Fermi surfaces on the boundary and the topological number in the bulk~\cite{Kitaev-classification, Ryu-classification, Teo-classification, FS-Classification, FS-TI}. In particular, from a general index theorem established by us~\cite{FS-TI}, we construct three exotic 1D models that have different types of topological charge of Majorana zero-modes and study rigorously the corresponding topological properties.


Let us first review briefly the bulk-boundary correspondence. We recall that a set of complete classification was obtained for strong topological insulators and superconductors(TIs/TSCs)  with the time-reversal symmetry(TRS), particle-hole symmetry(PHS), or/and chiral symmetry(CS), as illustrated in Tab.(\ref{tab:Periodic-Table}) in a right-to-left manner~\cite{Kitaev-classification, Ryu-classification, Teo-classification, FS-TI}. Then it is noted that  the complete classification of Fermi surfaces(FSs) was also achieved by considering the same set of symmetries~\cite{FS-Classification, FS-TI}, through generalizing a primary topological charge of Fermi surfaces~\cite{Volovik-Book, Volovik-Vacuum, Horava} as illustrated in Tab.(\ref{tab:Periodic-Table}) in a left-to-right manner, and the one-to-one relation between the classification of FSs and that of TIs/TSCs has also been clarified rigorously as a dimension-shift relation~\cite{FS-TI}. Moreover, a  general index theorem describing quantitatively the topological boundary-bulk correspondence of TIs/TSCs was established and proven~\cite{FS-TI, Note-index}, which is written as
\begin{equation}
\nu_{L,R} (d-1,i)=\pm N(d,i), \label{eq:Index-theorem}
\end{equation}
where the lefthand side denotes the total topological charge of FSs on the surface of a given TI/TSC and the righthand side is the topological number of the bulk. It is emphasised that the above general index theorem of Eq.(\ref{eq:Index-theorem}) also indicates the type of the topological charge of FSs is the same as that of the bulk topological number. Note that for a $\mathbf{Z}$-type TI/TSC, the topological charges of FSs at the left and right boundaries have the opposite signs, while for a $\mathbf{Z}_2$ type one, the sign of a topological charge plays no role.  For $d=1$, this general index theorem suggests that the models may have the Majorana zero-modes (FSs) with different topological origins from that of the Kitaev model, making it possible for us to have a new and deeper insight on all topological types of Majorana zero-modes at ends of topological superconducting wires, with exotic topological properties and phenomena as well as applications being expected.

\begin{table}
\begin{centering}
\begin{tabular}{|c|c|c|c|c|c|c|c|c|c|}
\hline
FS & AI & BDI & D & DIII & AII & CII & C & CI & $\frac{\mathrm{TI}}{\mathrm{TSC}}$\tabularnewline
\hline
\hline
T & +1 & +1 & 0 & -1 & -1 & -1 & 0 & +1 & T\tabularnewline
\hline
C & 0 & +1 & +1 & +1 & 0 & -1 & -1 & -1 & C\tabularnewline
\hline
S & 0 & 1 & 0 & 1 & 0 & 1 & 0 & 1 & S\tabularnewline
\hline
\hline
p\textbackslash{}i & 1 & 2 & 3 & 4 & 5 & 6 & 7 & 8 & i/d\tabularnewline
\hline
0 & 0 & $\mathbf{Z}$ & $\mathbf{Z}_{2}^{(1)}$ & $\mathbf{Z}_{2}^{(2)}$ & 0 & $2\mathbf{Z}$ & 0 & 0 & 1\tabularnewline
\hline
1 & 0 & 0 & $\mathbf{Z}$ & $\mathbf{Z}_{2}^{(1)}$ & $\mathbf{Z}_{2}^{(2)}$ & 0 & $2\mathbf{Z}$ & 0 & 2\tabularnewline
\hline
2 & 0 & 0 & 0 & $\mathbf{Z}$ & $\mathbf{Z}_{2}^{(1)}$ & $\mathbf{Z}_{2}^{(2)}$ & 0 & $2\mathbf{Z}$ & 3\tabularnewline
\hline
\end{tabular}
\par\end{centering}
\caption{Periodic Table of topological types for FSs~\cite{FS-Classification} and TIs/TSCs~\cite{Kitaev-classification, Ryu-classification, Teo-classification, FS-TI}.\label{tab:Periodic-Table} $d$,  $p$, and $i$ denote the spatial dimension of a TI/TSC, the codimension of an FS, and the index of symmetry classes, respectively. 
}
\end{table}

\section{Results}

At this stage, we are ready to focus on our propose, that is to investigate a series of special but very important 1D TIs/TSCs listed in the first row of Tab.(\ref{tab:Periodic-Table}), which is of both theoretical and practical interests.
It is seen clearly from Tab.(\ref{tab:Periodic-Table}) that for topologically nontrivial classes in one dimension consisting of classes BDI, D, DIII and CII, all have PHSs, which implies all nontrivial cases may be realized by TSCs, and therefore the gapless modes at ends may be topologically protected Majorana zero-modes with different topological types. Remarkably, through the general index theorem of Eq.(\ref{eq:Index-theorem}), we are now able to classify MFs according to the topological types of their nontrivial topological charges/numbers.
  We here construct the models for the other three nontrivial classes, rather than the class $D$ considered in the pioneer work of Kitaev~\cite{Kitaev}, and analyze their topological properties and the topological structures of Majorana zero-modes by using an analytical technique developed in~\cite{Kitaev}, under the unified picture provided by the general index theorem, Eq.(\ref{eq:Index-theorem}). Models in all symmetry classes can be constructed from a generic form in momentum space, $\mathcal{H}=\sum_{ij}f_{ij}(k)\sigma_i\otimes\tau_j$, where $f(k)$s are even or odd functions determined by the required symmetries and $\sigma$ and $\tau$ are Pauli matrices. Making comparisons with the Kitaev model, it is found that these three models, although being unified with the Kitaev model by the general index theorem, have their own special properties, most of which are essentially different from those of the Kitaev model because of their different topological essences. The special topology-related properties of each model may be utilized to universally store and manipulate quantum information beyond the existing models, as we will discuss case by case. In addition, we also note that although several models in classes DIII \cite{DIII-I,DIII-II, DIII-III, DIII-IV} and BDI \cite{BDI-I} have also been addressed, they are essentially reminiscent of the Kitaev model, without their own topological features being exposed, or are targeted to the realisation in concrete materials, rather than exploring the novel behaviours of the new Majorana zero-modes as what we have done here.

\textit{DIII-model}-
Let us begin with a model in class DIII, which has a TRS of minus sign and a PHS of positive sign, and thus may have a
$\mathbf{Z}_{2}^{(2)}$-type nontrivial topological number, as seen in Tab.(\ref{tab:Periodic-Table}). Recall that the Kitaev model is in class D with only a PHS of positive sign, and accordingly it may have a Majorana zero-mode at each end that protected by a $\mathbf{Z}_{2}^{(1)}$-type nontrivial topological number. According to our general index theorem, the Majorana zero-modes at the ends in class DIII should have a $ \mathbf{Z}_{2}^{(2)}$-type topological charge, in contrast to those of the Kitaev model having a $\mathbf{Z}_{2}^{(1)}$-type topological charge. Just like that of higher dimensions where the two $\mathbf{Z}_{2}$-type topological charges correspond to different configurations of Fermi surfaces~\cite{FS-TI,FS-Classification}, in contrast to that they are not distinguished in earlier works~\cite{Kitaev-classification,Ryu-classification,Teo-classification}, it is expected that the Majorana zero-modes in the present DIII model are essentially different from those in the Kitaev model. To be explicit, the unique features of our DIII model are listed as followings: (\textit{i}) Our DIII model in its topological phase has, at each end, two Majorana zero-modes of orthogonal spin pairings as the eigen ones of $\sigma_x$, being different from that in the Kitaev model where only one spinless Majorana zero-mode resides at each end; (\textit{ii}) The ground-state degeneracy is four, two times of the ground-state degeneracy of the Kitaev model; (\textit{iii}) The relatively rich diversity of this DIII model creates more possibilities for quantum manipulation, such as quantum information storage and processing. Most significantly, this DIII model with one Majorana fermion being generated at each end is in a two-fold degenerate space, which may be utilized as a physical qubit for universal quantum manipulation, namely an arbitrarily given state can be created by coupling the system to a weak effective magnetic field. In contrast, only one quantum gate can be realized with a conventional method using the two copies of Kitaev modes through complicated braiding operations~\cite{Exchange-MFs,Review-MFs}. Furthermore, there exists no difficulty in principle that a universal quantum computation can be realized with additional braiding operations acting on two copies of our DIII model, while for the Kitaev model this is in principle impossible.

\begin{figure}
\begin{centering}
\includegraphics[scale=0.40]{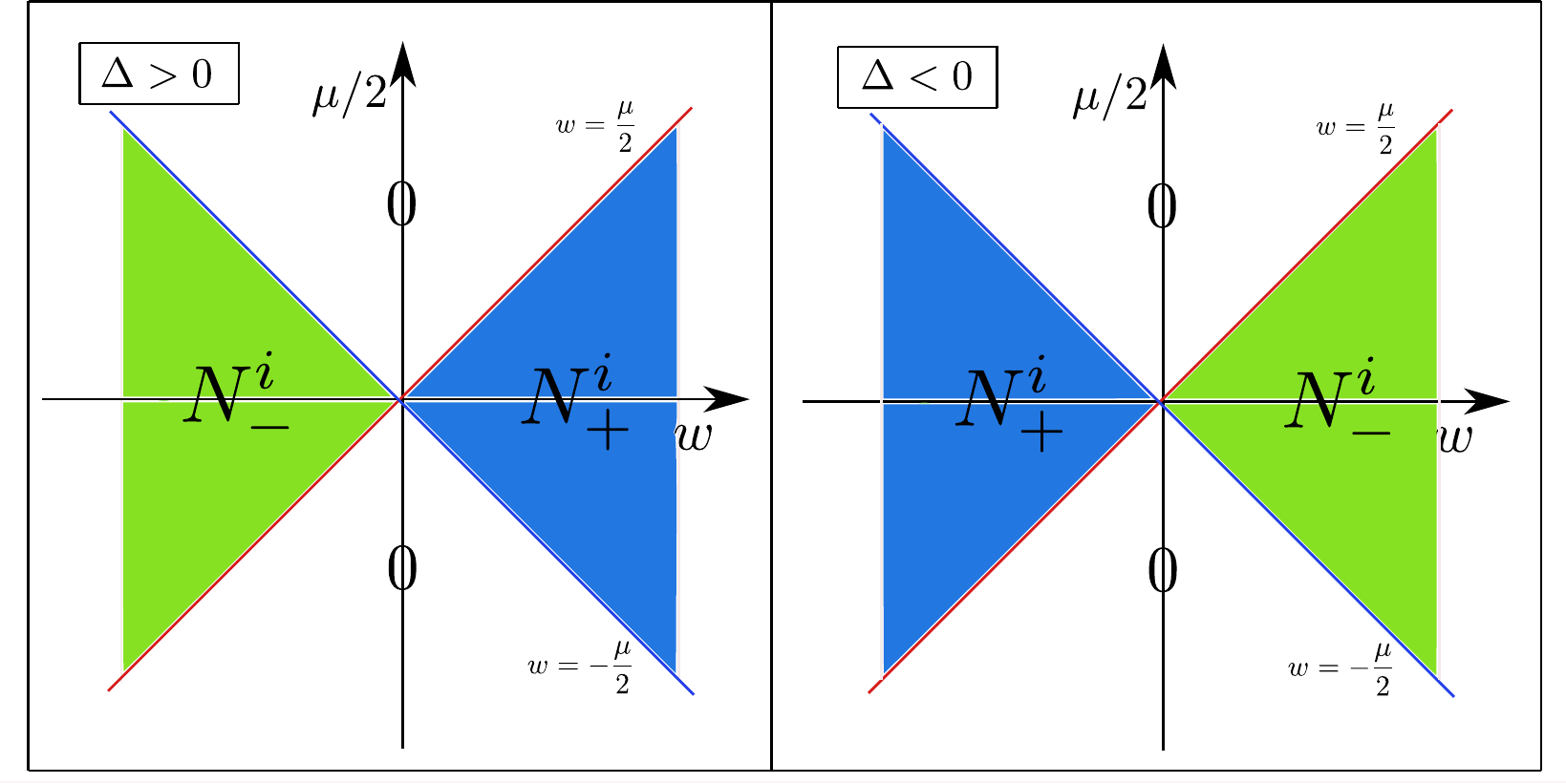}
\par
\end{centering}
\caption{Topological Phase Diagram. $N^i_{w>/<0}$ denotes the topological number of each model in the $i$th class when $w>/<0$ in a nontrivial region. Specifically, $N^{D}_{w>0}=N^{D}_{w<0}=1\mod 2$, $N^{BDI}_{w>0}=-N^{BDI}_{w<0}=1$, $N^{DIII}_{w>0}=N^{DIII}_{w<0}=1\mod 2$, and $N^{CII}_{w>0}=-N^{CII}_{w<0}=-2$.\label{fig:Phase-Diagram}}
\end{figure}

 We now start to introduce our DIII model, whose Hamiltonian reads
\begin{eqnarray}
H_{DIII} & = & \sum_{j}\left[-w\left(a_{j}^{\dagger}a_{j+1}+a_{j+1}^{\dagger}a_{j}\right)-\mu\left(a_{j}^{\dagger}a_{j}-1\right)\right]\nonumber \\
 &  & -\Delta\sum_{j}\left(a_{j+1}i\sigma_{1}a_{j}+a_{j}^{\dagger}(-i\sigma_{1})a_{j+1}^{\dagger}\right)\nonumber \\
 & = & \int dk\:\Psi_{k}^{\dagger}((-w\cos k-\mu/2)\mathbf{1}\otimes\tau_{3}\nonumber\\
 &  &\,\,\,\,\,\,\, -\Delta\sin k\sigma_{3}\otimes\tau_{1})\Psi_{k},
\label{eq:DIII-Model}\end{eqnarray}
where $w$, $\mu$ and $\Delta$ are all real,  $\sigma_i$ and $\tau_i$ are Pauli matrices, $\Psi_{k}^{\dagger}=(a_{k}^{\dagger},\: a_{-k}(-i\sigma_{2}))$, and the spin index is implicit. From the expression of $\mathcal{H}_{DIII}(k)$ after the second equality in Eq.(\ref{eq:DIII-Model}), it is seen that model has a PHS with $C=\tau_1$ and a TRS with $T=\sigma_2$ \cite{Note-symmetry}, which confirm that it is in class DIII. The model may be viewed as a spin-triplet BdG model. According to Tab.(\ref{tab:Periodic-Table}), it may have a $\mathbf{Z}_2^{(2)}$-type topological number. Postponing the calculation details in Methods, the regime of $|w|>|\mu/2|$ is identified as the topological phase and the phase diagram of this model is shown in Fig.(\ref{fig:Phase-Diagram}). The non-trivial topological number implies Majorana zero-modes of $\mathbf{Z}_2^{(2)}$-type topological charge at each end from our general index theorem Eq.(\ref{eq:Index-theorem}). So we are now going to look into  these Majorana zero-modes in detail.

\begin{figure}
\begin{centering}
\includegraphics[scale=0.28]{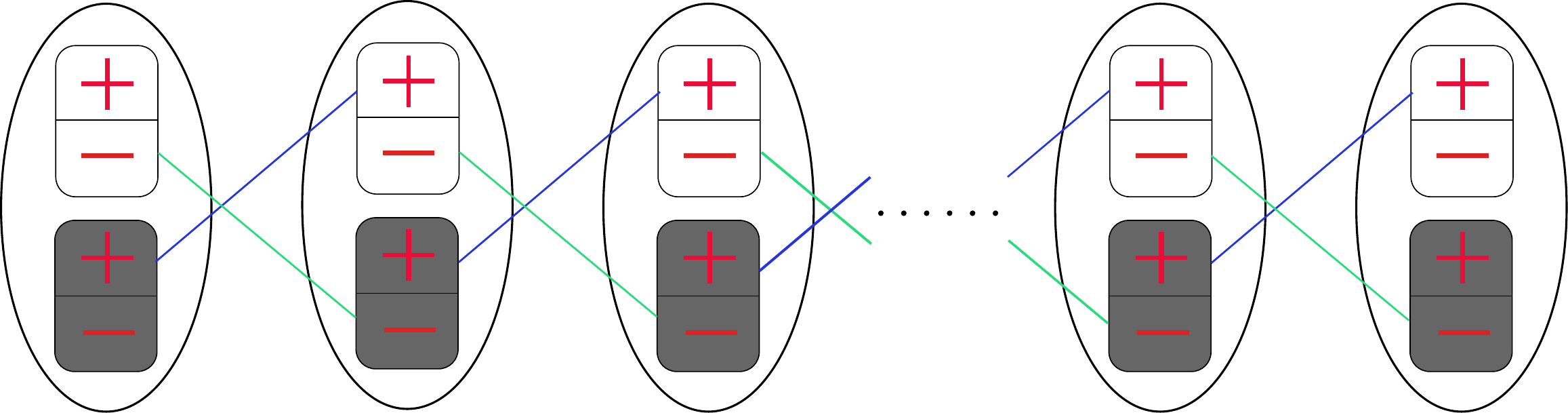}
\par\end{centering}
\caption{The Majorana repairing of the DIII model.\label{fig:DIII-pairing}}
\end{figure}

As usual,   the  Majorana operators  can be decomposed as $\gamma_{js}^{1}=e^{i\frac{\theta}{2}}a_{js}+e^{-i\frac{\theta}{2}}a_{js}^{\dagger}$ and $\gamma_{js}^{2}=(e^{i\frac{\theta}{2}}a_{js}-e^{-i\frac{\theta}{2}}a_{js}^{\dagger})/i$,
where $\gamma^{a}$ with $a=1,2$ satisfy $\gamma^\dagger=\gamma$, and $s$ is the spin index. It is pinpointed that $\theta$ in the above decomposition can be chosen arbitrarily, and is physically indistinguishable in our DIII model, in contrast to that the $\theta$ in the Kitaev mode is related to the phase of the cooper-pairing order-parameter.  Since the existence of unpaired Majorana zero-modes and their features are topologically intrinsic according to the general index theorem, Eq.(\ref{eq:Index-theorem}), to see the structure of these zero-modes it is sufficient to solve the model exactly at one specific point in the topological phase, which is one of main contributions from Kitaev~\cite{Kitaev, Kitaev-Honeycomb}.  For simplicity, adopting the convention that $\theta=\pi/2$, let us focus on the case $w=\Delta$ and $\mu=0$ in the topological phase, where our DIII model (\ref{eq:DIII-Model}) is reduced to
$H_{DIII}=\sum_{j}iw\gamma_{j,+}^{2}\gamma_{j+1,+}^{1}+\sum_{j}-iw\gamma_{j,-}^{1}\gamma_{j+1,-}^{2}, \label{eq:Simplified-DIII}$
where `$\pm$' are two eigen spin pairings with respect to $\sigma_x$, i.e.,
$a_{j,+}=\frac{1}{\sqrt{2}}(a_{j,\uparrow}+a_{j,\downarrow})$ and
$a_{j,-}=\frac{1}{\sqrt{2}}(a_{j,\uparrow}-a_{j,\downarrow})$.
It is observed that $\gamma^2_{j,\alpha}$ and $\gamma^1_{j+1,\alpha}$ (with $\alpha=\pm$) can pair together to form new electronic operators, $\tilde{a}_{j,+}=\frac{1}{2}(\gamma_{j,+}^{2}+i\gamma_{j+1,+}^{1})$ and $\tilde{a}_{j,-}=\frac{1}{2}(\gamma_{j+1,-}^{2}+i\gamma_{j,-}^{1})$, to fully diagonalize the Hamiltonian. The pairing pattern of Majorana operators under the open-boundary condition is pictured in Fig.(\ref{fig:DIII-pairing}). It is observed that  $\gamma^1_{1,+}$ and $\gamma^2_{N,+}$(with N being the total number of sites) are unpaired without entering the Hamiltonian at the two ends, and so are $\gamma^2_{1,-}$ and $\gamma^1_{N,-}$. In other words,   $\gamma^1_{1,+}$ and $\gamma^2_{1,-}$ are Majorana zero-modes at the left end, while $\gamma^2_{N,+}$ and $\gamma^1_{N,-}$ are the ones  at the right end. To compare with the Kitaev model of $\mathbf{Z}_2^{(1)}$-type, recall that only one Majorana operator is unpaired at each end in its topological phase.

 It is illuminating to view these Majorana zero-modes according to the general index theorem (\ref{eq:Index-theorem}). It is observed that  $\gamma^1_{1,+}$ and $\gamma^2_{1,-}$ together have nontrivial $\mathbf{Z}_2^{(2)}$-type topological charge $\nu_L=1$ at the left end , meanwhile  $\gamma^2_{N,+}$ and $\gamma^1_{N,-}$  also have $\nu_R=1$ at the right end. Since Majorana zero-modes at the two ends have the same topological charge, they should be able to be deformed to each other continuously. As $\theta$ in our DIII model can be arbitrarily chosen, it is seen that  $\gamma_j^1$ is transformed to $\gamma_j^2$ after rotating $\theta$ by $\pi$ in the general Majorana decomposition, confirming that Majorana zero-modes at each end are indeed topologically equivalent in agreement with their $\mathbf{Z}_2$ nature. For comparison with the $\mathbf{Z}_2^{(1)}$-type Kitaev model,  $\theta$ in the present DIII model is an intrinsic freedom such that varying it keeps the Hamiltonian invariant, while its variation in the Kitaev mode changes the Hamiltonian explicitly in a topological equivalent family. 

This result is, although obtained at a specific point in the topological phase, can be generalized to a generic situation according to the topological protection. The Majorana zero-mode represented by  $\gamma^1_{1,+}$, for instance, is replaced by $\Gamma^L_+$ as a linear combination of Majorana operators with real coefficients and  `$+$' spin pairing, and the other three Majorana zero-modes are replaced by $\Gamma^L_-$, $\Gamma^R_+$ and $\Gamma^R_-$, respectively, in the same sense.  For all of the four combinations, their coefficients decay exponentially away from the ends 
in the thermodynamic limit $N\longrightarrow \infty$. Because four Majorana operators do not enter the Hamiltonian, the ground state are four-fold degenerate with a basis as
$\{|\psi^+_1\psi^-_1\rangle,\;|\psi^+_0\psi^-_1\rangle,\;|\psi^+_1\psi^-_0\rangle, \;|\psi^+_0\psi^-_0\rangle\}$.
The first one has no Majorana fermion, the second and the third one have one Majorana fermion at each end, and the last one has two Majorana fermions at each end. Since  $-i\Gamma^{L}_{\alpha}\Gamma^{R}_{\alpha}|\psi_{0,1}^{\alpha}\rangle=\pm|\psi_{0,1}^{\alpha}\rangle$, it is straightforward  to see that the states of no Majorana fermion and of two Majorana fermions at each end have the even fermionic parity, i.e. the number of fermions is even, and the two states with one Majorana at each end have the odd fermionic parity. Also for comparison, we recall that the ground state of the Kitaev model with one Majorana fermion at each end has the even fermionic parity, while the ground state without Majorana fermions has the odd fermionic parity~\cite{Kitaev}.

The ground space, spanned by  $|\bullet\rangle=|\psi^+_0\psi^-_1\rangle$ and$|\times\rangle=|\psi^+_1\psi^-_0\rangle$ with one Majorana fermion at each end, has odd parity distinguishing it from the other ground states. Because of the orthogonal spin-pairings for $|\bullet\rangle$ and $|\times\rangle$, the degeneracy may be broken by coupling the spin degree of the DIII model to an external source, and therefore the space can be regarded as a qubit whose any state is achievable. Let us demonstrate this by coupling all electron spins to a very weak external magnetic field compared with the bulk energy gap, i.e. $H'=\sum_{j}1/2\mathbf{B}\cdot\sigma_j$ as perturbation. After detailed calculations in the electronic representation (see the section of Methods), we have the low-energy effective results: $\langle\bullet|\hat{S}_x|\bullet\rangle=1/2$, $\langle\times|\hat{S}_x|\times\rangle=-1/2$, $\langle\bullet|\hat{S}_{y,z}|\bullet\rangle=0$, and $\langle\times|\hat{S}_{y,z}|\times\rangle=0$ under the thermodynamic limit, with $\hat{\mathbf{S}}=\sum_j1/2\sigma_j$. An intuitive picture of this feature is obviously that the MF states are wave pockets entering at the ends with the specific spin orientations being the only degrees that are sensitive to the external magnetic field, irrelevant to the homogeneous bulk state. Thus as what the symbols suggest, $\hat{\mathbf{S}}$ may be effectively regarded as a single spin operator acting on the two-level space with $|\bullet\rangle$ and $|\times\rangle$ being the two eigenstates of $\hat{S}_x$. Thus through a suitable control of $H'(t)$ with $H'(t=0)=H'(t=t_f)=0$, an arbitrary state can be created, leading to a universal single-qubit operation.

\textit{BDI-model}-
According to Tab.(\ref{tab:Periodic-Table}), in class BDI with a TRS of minus sign and a PHS of plus sign, there may be $\mathbf{Z}$-type nontrivial  systems. In this section we introduce a nontrivial model of simple form, which shows essential differences between $\mathbf{Z}$-type systems and $\mathbf{Z}_2^{(1,2)}$-type ones. The BDI model has two nontrivial topological phases corresponding to the bulk topological number $N=\pm1$, in contrast to $\mathbf{Z}_2^{(1,2)}$-types, for which $1=-1\mod 2$ in the same topological phase. In the phase of $N=1$, according to our general index theorem (\ref{eq:Index-theorem}), the model has a Majorana zero-mode at its left end that has topological charge $\nu_L=1$, and the other Majorana zero-mode at its right end that has topological charge $\nu_R=-1$. On the other hand, in the phase of $N=-1$, reversely, $\nu_L=-1$ and $\nu_R=1$. These predictions from the general index theorem (\ref{eq:Index-theorem}) are confirmed by our BDI model, where the topological charges of Majorana zero-modes $\nu=\pm1$ are identified as the two distinct Majorana operators decomposed from an electronic physical freedom, just like those in DIII case without spin degree. We pinpoint that the two phases are also distinguished by the fact that their ground states are orthogonal to each other. It is interesting to discuss the phase transition between the two phases $N=\pm1$. Firstly, after the transition from $N=1$ to $N=-1$, the unpaired Majorana operators at the two end are exchanged showing the exchange of topological charges of Majorana zero-modes. Secondly, there is a novel way to realize the phase transition without closing the spectrum gap, but through breaking the TRS, so that the two phases are interpolated by a series of models in class D with nontrivial $\mathbf{Z}_2^{(1)}$-type topological number.  In principle, we can construct BDI models with large topological number $N$, and because it belongs to $\mathbf{Z}$-type, there are $N$ unpaired Majorana operators at each end without entering the Hamiltonian, corresponding to $2^N$-fold ground state degeneracy. These exotic features provide more possibilities to store and process quantum information, and might be utilized to realize topological quantum computing.

 The Hamiltonian of the BDI model reads
\begin{eqnarray}
H_{BDI} & = & \sum_{j}\left(-wa_{j}^{\dagger}a_{j+1}+i\Delta a_{j}a_{j+1}\right)+\mathrm{h.c.}\nonumber\\
 &  & \qquad-\mu(a_{j}^{\dagger}a_{j}-\frac{1}{2}) \label{eq:BDI-Model}\\
 & = & \int dk\;\Psi_{k}^{\dagger}(-\Delta\sin k\tau_{1}-(w\cos k+\mu/2)\tau_{3})\Psi_{k}, \nonumber
\end{eqnarray}
where $\Delta$ is still real, $a_j$ are spinless, and therefore $\Psi_{k}=(a_k,\,\,a_{-k}^\dagger)^T$. From the second equality, we read $\mathcal{H}_{BDI}(k)$, which has a TRS with $T=\tau_3$ and a PHS with $C=\tau_2$ corresponding to class BDI. According to Tab.(\ref{tab:Periodic-Table}), the formula for $\mathbf{Z}$-type topological number is used to calculate the topological number in Methods, and the resulted phase diagram is shown in Fig.(\ref{fig:Phase-Diagram}). In topological phases, i.e. $|w|>|\mu/2|$, there exist Majorana zero-modes at two ends, noting that now $N=\pm1$ are distinguishable corresponding to different topological phases. The Majorana decompositions are adopted as
$\gamma_{j}^{1}=e^{i\frac{\theta}{2}}a_{j}+e^{-i\frac{\theta}{2}}a_{j}^{\dagger}$ and
$\gamma_{j}^{2}=(e^{i\frac{\theta}{2}}a_{j}-e^{-i\frac{\theta}{2}}a_{j}^{\dagger})/i$.

We start from investigating the Majorana zero-modes when $N=+1$, and the representative point in the topological phase is chosen as $w=\Delta>0$ and $\mu=0$, which is exactly solvable. Through fixing $\theta=\pi/2$ in the Majorana decompositions, we reduce the Hamiltonian as $H_{BDI} = \sum_{j}iw \gamma_{j}^{2}\gamma_{j+1}^{1}\label{eq:Simplified-BDI}$, which can be fully diagonalized by using the repaired electronic operator $\tilde{a}_j=\frac{1}{2}(\gamma_{j}^{2}+i\gamma_{j+1}^{1})$. It is observed that $\gamma^1_1$ and $\gamma^2_N$ do not enter the Hamiltonian, serving as Majorana zero-modes at two ends.  According to the general index theorem (\ref{eq:Index-theorem}), we assign $\mathbf{Z}$-type topological charge $\nu_L[c'^1_1]=+1$ and $\nu_R[c'^2_N]=-1$ for the two zero-modes at two ends, respectively, with `$\prime$' indicating that $\theta=\pi/2$ in the Majorana decompositions.

The simple case corresponding to the topological phase with $N=-1$ is $w=-\Delta>0$ and $\mu=0$. We can still simplify the Hamiltonian by using the Majorana decompositions,  but in this case fixing $\theta=-\pi/2$. In this phase, similarly $\gamma''^1_1$ and $\gamma''^2_N$ are Majorana zero-modes at two ends with `$\prime\prime$' noting that $\theta=-\pi/2$ different from that in the phase $N=+1$. The general index theorem (\ref{eq:Index-theorem}) implies that $\nu_L[\gamma''^1_1]=-1$ and $\nu_R[\gamma''^2_N]=1$. It is direct to check that $\gamma''^1_j=\gamma'^{2}_{j}$ and $\gamma''^2_j=-\gamma'^{1}_{j}$, which conforms to our general index theorem.

\begin{figure}
\includegraphics[scale=0.35]{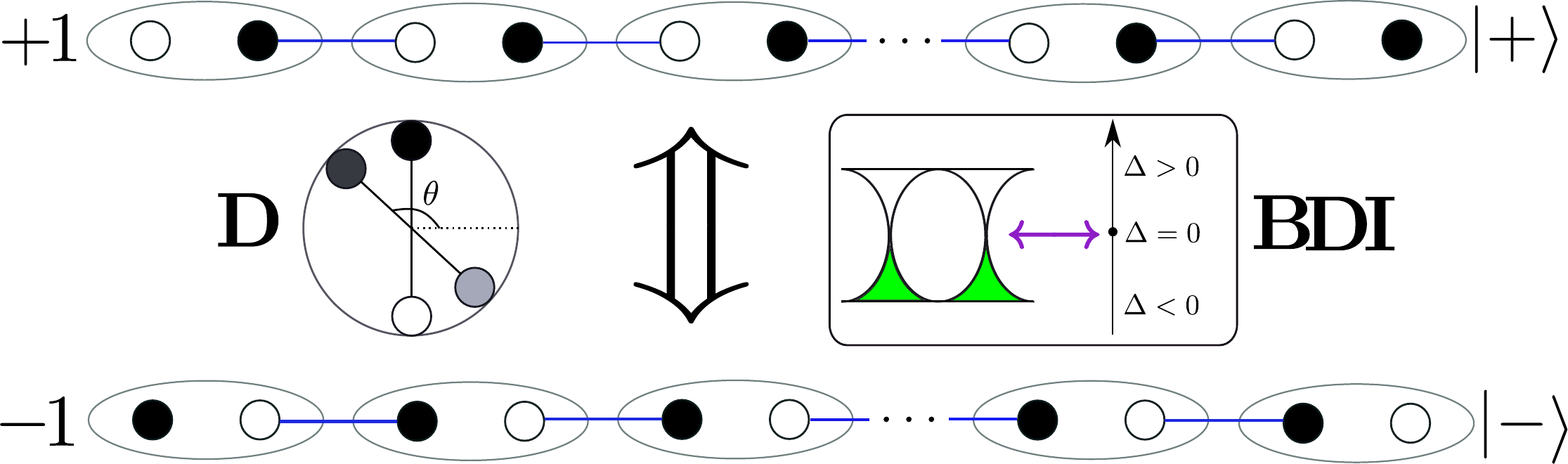}
\caption{Phase transition of the BDI model. \label{fig:BDI-Transition}}
\end{figure}

We proceed to address the topological phase transition between $N=\pm1$ in our BDI model. If the Hamiltonian is restricted to be in class BDI, the phase transition always implies gap closing in the bulk spectrum, because the two phases corresponds to distinct topological classes. This can be realized through, for instance, continuously varying $\Delta$ or $w$ from being positive to negative, where $\Delta=0$ or $w=0$ corresponds to transition point with gap closing. However, there exists another way to realize the topological phase transition through breaking the TRS, an anti-unitary symmetry, without closing the gap in the bulk, which is essentially different from ordinary phase transitions breaking a unitary continuous symmetry (O(3) in a magnet for instance) in Landau's paradigm. We can break the TRS with $T=\tau_3$ to make the system in class D by varying $\Delta$ to be complex, namely replacing $i\Delta$ in Eq.(\ref{eq:BDI-Model}) by $|\Delta|e^{i\theta}$. Then $\theta=\pm\pi/2$ correspond to the two topological phases of $N=\pm1$, respectively, and the two phases can be connected by continuously varying $\theta$ from $\pi/2$ to $-\pi/2$ with $|\Delta|$ fixed. The two ways for phase transition are illustrated in Fig.(\ref{fig:BDI-Transition}). Furthermore, we pinpoint that during the process of varying $\theta$, the system is protected by a nontrivial $\mathbf{Z}_2^{(1)}$-type topology, while only when $\theta=\pm\pi/2$ the $\mathbf{Z}$-type topological protection is recovered, noting that when $\Delta$ is complex, our BDI model is just the Kitaev model~\cite{Kitaev}. To see the distinction of the two phases clearly, it is noted that the ground states of the two phases under open-boundary condition are orthogonal to each other in the thermodynamic limit (see Methods), which is expressed as $\langle \psi_\alpha,1|\psi_\beta,-1\rangle=0$ with $|\psi_\alpha,\pm1\rangle$ being a generic ground state for $N=\pm1$.

\begin{figure}
\begin{centering}
\includegraphics[scale=0.28]{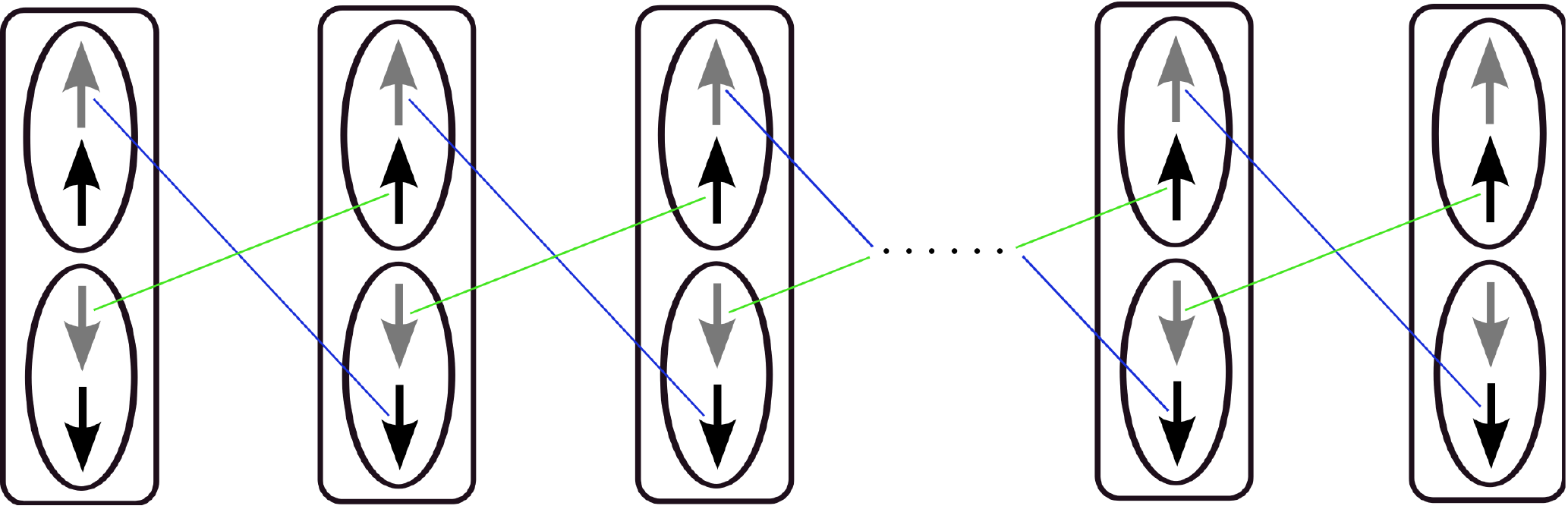}
\par\end{centering}
\caption{The Majorana repairings of the CII model.\label{fig:CII-pairing}}
\end{figure}

 \textit{CII model}-
Let us introduce an exotic model in class CII, which  has $\mathbf{Z}$-type topological number $N=\pm2$ in its topological phases. With the experience from the two models already demonstrated, we just brief the main points. The Hamiltonian is
\begin{eqnarray}
H_{CII} & = & \sum_{j}(-\Delta a_{j+1}^{\dagger}i\sigma_{2}a_{j}+w a_{j+1}^{\dagger}i\sigma_{2}a_{j}^{\dagger}\nonumber\\
 &  & \qquad-\mu a_{j}^{\dagger}i\sigma_{2}a_{j}^{\dagger})+\mathrm{h.c.}\nonumber\\
 & = & \int dk\:\Psi_{k}^{\dagger}(-\Delta\sin k\sigma_{2}\otimes\tau_{3}\nonumber\\
 &  & \qquad+(w\cos k-\mu)\mathbf{1}\otimes\tau_{1})\Psi_{k},\label{eq:CII-Model}
\end{eqnarray}
where still $\Delta$ and $w$ are real, and $\Psi_k$ is the same as that of the DIII model (\ref{eq:DIII-Model}). The $\mathcal{H}_{CII}(k)$ read from the second equality has a TRS with $T=\sigma_2\otimes\mathbf{1}$ and a PHS with $C=\mathbf{1}\otimes\tau_2$. Although the exotic form of the Hamiltonian with special spin-pairings, it may be simulated by various artificial systems. Straightforward calculation using the corresponding $\mathbf{Z}$-type formula shows it has topological number $N=\pm2$ when $|w|>|\mu/2|$ as in Fig(\ref{fig:Phase-Diagram}), consistent with $2\mathbf{Z}$ in Tab.(\ref{tab:Periodic-Table}). To see the form of the topologically protected Majorana zero-modes at two ends, when $w=\Delta>0$ and $\mu=0$ corresponding to $N=2$, we adopt the Majorana decompositions of the DIII model with $\theta=0$, and simplify the CII model (\ref{eq:CII-Model}) as
$H_{CII}=\sum_{j}iw\gamma_{j\downarrow}^{2}\gamma_{j+1\uparrow}^{1}-\sum_{j}iw\gamma_{j\uparrow}^{2}\gamma_{j+1\downarrow}^{1}$, which may be viewed as two copies of the simplified BDI model. The Hamiltonian can be fully diagonalized by introducing $\tilde{a}_j=\frac{1}{2}(\gamma^2_{j\downarrow}+i\gamma^1_{j+1\uparrow})$ and $\tilde{b}_j=\frac{1}{2}(\gamma^2_{j\uparrow}+i\gamma^1_{j+1\downarrow})$, which is illustrated in Fig.(\ref{fig:CII-pairing}). For the left end, $\gamma^1_{1\uparrow}$ and $\gamma^1_{1\downarrow}$ serve as Majorana zero-modes, each having $\mathbf{Z}$-type topological charge $\nu_L[\gamma^1_{\uparrow,\downarrow}]=1$, and accordingly for the right end $\nu_R[\gamma^2_{\uparrow,\downarrow}]=-1$, which follows the general index theorem (\ref{eq:Index-theorem}). Consistent topological charges of Majorana zero-modes can be assigned for the other topological phase $N=-2$.  Note that the explanation of the CII model by the general index theorem is entirely parallel to that of the BDI model, since they both belong to $\mathbf{Z}$-type. Although similar to two copies of the BDI model, however we underline that there does not exist a way to connect the two topological phases of the CII model, 
since breaking either the TRS or PHS makes the topology trivial according to Tab.(\ref{tab:Periodic-Table}).

\section{Methods}
\textit{Topological numbers}-
The DIII model is of the $\mathbf{Z}_2^{(2)}$ topological type, and therefore a two-parameter symmetry-preserving extension of $\mathcal{H}_{DIII}$, Eq.(\ref{eq:DIII-Model}),  should be made below to calculate its topological number~\cite{FS-TI}:
\begin{eqnarray*}
\mathcal{H}_{DIII}(k,\theta,\phi) & = & \left[\mathcal{H}(k)\cos\theta+\sin\theta\mathbf{1}\otimes\tau_{2}\right]\cos\phi\\
 & &+\sin\phi\sigma_{1}\otimes\tau_{1}\\
 & = & (w\cos k+\mu)\cos\theta\cos\phi\mathbf{1}\otimes\tau_{3}\\
 &  & +\sin k\cos\theta\cos\phi\sigma_{3}\otimes\tau_{1}\\
 &  &+\sin\theta\cos\phi\mathbf{1}\otimes\tau_{2}+\sin\phi\sigma_{1}\otimes\tau_{1}.
\end{eqnarray*}
In the calculation, the following formula has been used:
\begin{eqnarray*}
N_{DIII}& = & \frac{1}{48\pi^{2}}\int_{0}^{2\pi}dk\int_{-\frac{\pi}{2}}^{\frac{\pi}{2}}d\theta\int_{-\frac{\pi}{2}}^{\frac{\pi}{2}}d\phi\:\epsilon^{\mu\nu\lambda}\\
& &\,\,\,\times\mathbf{tr}\left[\sigma_{2}\otimes\tau_{1}\mathcal{H}^{-1}\partial_{\mu}\mathcal{H}\mathcal{H}^{-1}\partial_{\nu}\mathcal{H}\mathcal{H}^{-1}\partial_{\lambda}\mathcal{H}\right].
\end{eqnarray*}
The BDI model and the CII model are both belong to the $\mathbf{Z}$ type, and therefore their corresponding formulae are respectively~\cite{FS-TI}
\[
N_{BDI}=\frac{1}{4\pi i}\int_{-\pi}^{\pi}dk\:\mathbf{tr}\left(\tau_{2}\mathcal{H}^{-1}(k)\partial_{k}\mathcal{H}(k)\right).
\]
and
\[
N_{CII}=\frac{1}{4\pi i}\int_{-\pi}^{\pi}dk\:\mathbf{tr}\left(\sigma_{2}\otimes\tau_{2}\mathcal{H}^{-1}(k)\partial_{k}\mathcal{H}(k)\right).
\]
The results obtained from the above three formulae are illustrated in the phase diagram, Fig(\ref{fig:Phase-Diagram}).

\textit{Orthogonality in BDI}-
We first solve the ground states of the BDI model exactly in the electronic representation when $w=|\Delta|>0$ and $\mu$ for a generic $\theta$ and then compare the ground states of $\theta=\pi/2$ and $\theta=-\pi/2$.  The explicit expressions of the two degenerate ground states are
\begin{equation*}
|\psi_0\rangle=\sum_{n=0}^{[N/2]}\sum_{{\scriptscriptstyle s_{1}<\cdots<s_{2n}}}e^{-in\theta}a_{s_{1}}^{\dagger}a_{s_{2}}^{\dagger}\cdots a_{s_{2n}}^{\dagger}|0\rangle
\end{equation*}
and
\begin{equation*}
|\psi_1\rangle=\sum_{n=0}^{[(N-1)/2]}\sum_{{\scriptscriptstyle s_{1}<\cdots <s_{2n+1}}}e^{-i(n-1)\theta}a_{s_{1}}^{\dagger}a_{s_{2}}^{\dagger}\cdots a_{s_{2n+1}}^{\dagger}|0\rangle,
\end{equation*}
where `$[r]$' is the integer part of the real number $r$,  $|\psi_0\rangle$ with even parity has a pair of MFs at the two ends and $|\psi_1\rangle$ with odd parity has no MFs. The inner products of ground states for $N=\pm1$ are computed as
\begin{equation*}
\langle \psi_0,1|\psi_0,-1\rangle=\frac{C_{N}^{0}-C_{N}^{2}+C_{N}^{4}-\cdots}{C_{N}^{0}+C_{N}^{2}+C_{N}^{4}+\cdots}\longrightarrow0
\end{equation*}
and
\begin{equation*}
\langle\psi_1,1|\psi_1,-1\rangle=\frac{C_{N}^{1}-C_{N}^{3}+C_{N}^{5}-\cdots}{C_{N}^{1}+C_{N}^{3}+C_{N}^{5}+\cdots}\longrightarrow0,
\end{equation*}
where `$\longrightarrow$' denotes thermodynamic limit $N\rightarrow \infty$. It is obvious that $\langle \psi_1|\psi_0\rangle=0$.

\textit{Response of DIII}-
Using the above expressions of $|\psi_0\rangle$ and $|\psi_1\rangle$, the explicit expressions of the four degenerate ground states of the DIII model can also be constructed. Thus every element of the perturbation matrix $\hat{\mathbf{S}}$ can be computed directly. Here we only present the most important steps:
\begin{eqnarray*}
\delta E/B & = & \langle\psi_{0}^{-}\psi_{1}^{+}|\sum_{j}\sigma_{j}^{1}|\psi_{0}^{-}\psi_{1}^{+}\rangle-\langle\psi_{0}^{+}\psi_{1}^{-}|\sum_{j}\sigma_{j}^{1}|\psi_{0}^{+}\psi_{1}^{+}\rangle\\
 & = & \frac{1}{\sum_{\{m=0,n=0\}}^{\{[(L-1)/2],[L/2]\}}C_{L}^{2m+1}C_{L}^{2n}}\\
 & &\bigg[\sum_{m}^{[L/2]}\sum_{n}^{[(L-1)/2]}(-2m+2n+1)C_{L}^{2m}C_{L}^{2n+1}\\
 &  & -\sum_{n}^{[L/2]}\sum_{m}^{[(L-1)/2]}(-2m-1+2n)C_{L}^{2m+1}C_{L}^{2n}\bigg]\\
 & = & 2+\frac{4\sum_{m}^{[L/2]}\sum_{n}^{[(L-1)/2]}(n-m)C_{L}^{2m}C_{L}^{2n+1}}{\sum_{\{m=0,n=0\}}^{\{[(L-1)/2],[L/2]\}}C_{L}^{2m+1}C_{L}^{2n}}
\end{eqnarray*}
In the thermodynamic limit, the second term in the last equality vanishes, and thus we have $\delta E/B=2$. This justifies $\langle\bullet|\hat{S}_x|\bullet\rangle=1/2$ and $\langle\times|\hat{S}_x|\times\rangle=-1/2$, and the other two matrix elements can be obtained similarly.


{\it Acknowledgments}
We thank J. Teo and S. L. Zhu for helpful discussions. This work was supported
by the GRF (HKU7058/11P\&HKU7045/13P),  the CRF (HKU8/11G) of Hong Kong, and the URC fund of HKU.


\begin{thebibliography}{References}
\bibitem{Note-index} Strictly speaking, the index theorem for 1D TIs/TSCs is not explicitly characterized in Ref.\cite{FS-TI}, since topological charges for point-like gapless modes are not formulated in Ref.\cite{FS-Classification}, though it seems to be indicated by the eight-fold periodicity addressed there. Thus the present work actually completes the index theorem by explicitly revealing its meaning in 1D TIs/TSCs.
\bibitem{Note-symmetry}In our convention,  the anti-unitary operators of TRS and THS  are respectively represented as $\hat{T}=T\hat{\kappa}$ and $\hat{C}=C\hat{\kappa}$,  where $T$ and $C$ are linear unitary operators,  and $\hat{\kappa}$ denotes the basic anti-unitary operator of complex conjugation.
\end{thebibliography}
\end{document}